\documentclass[reprint,amsmath,amssymb,aps,prb]{revtex4-2}

\usepackage{graphicx}
\usepackage{epstopdf}
\usepackage{amsmath,bm}
\usepackage{threeparttable}
\usepackage{multirow}
\usepackage{booktabs}
\usepackage{color}
\usepackage{hyperref}
\hypersetup{colorlinks=true,linkcolor=blue,filecolor=blue,urlcolor=blue,citecolor=blue,
	}

\begin{document}

\title{High-Order Topological Phase Diagram Revealed by Anomalous Nernst Effect in Janus ScClI Monolayer}%
\author{Ning-Jing Yang$^{1,2}$}%
\author{Jian-Min Zhang\thanks{awaisadnan@gmail.com}$^{1,2}$}%
\email[Corresponding author:]{ jmzhang@fjnu.edu.cn} 

\affiliation{1 Fujian Provincial Key Laboratory of Quantum Manipulation and New Energy Materials, College of Physics and Energy, Fujian Normal University, Fuzhou 350117, China
}
\affiliation{2 Fujian Provincial Collaborative Innovation Center for Advanced High-Field Superconducting Materials and Engineering, Fuzhou, 350117, China
}

\begin{abstract}
Higher-order topological properties of two-dimensional(2D) magnetic materials have recently been proposed. 
In 2D ferromagnetic Janus materials, we find that ScClI is a second-order topological insulator (SOTI). Using the tight-binding approximation, we develop a multi-orbital model that adequately describes the high-order topological states of ScClI. Further, we give the complete high-order topological phase diagram of ScClI, based on the external field modulation of the magneto-valley coupling and energy levels. 
2D ScClI has a pronounced valley polarization. This type of magneto-valleytronics results in  different insulating phases to exhibit completely different anomalous Nernst conductance. 
Through valleytronics, we establish a link between the topological insulator and the valley Nernst effect, thus constructing an anomalous valley nernst conductance map that corresponds to the topological phase diagram.
We utilize the characteristics of valley electronics to link higher-order topological materials with the anomalous Nernst effect, which has potential implications for high-order topological insulators and valley electronics.
\end{abstract}

\maketitle

\section{Introduction}
Recently, two-dimensional high-order topological insulator(HOTI) have received great attention \cite{1,2,3,4,5,5_1}. Prior to this, higher order topological states in three-dimensional $\rm Bi,\ MnBi_{2n}Te_{3n+1}$, and $\rm EuIn_2As_2$ have received attention \cite{6,7,8,9,9_1}. However, higher-order topological corner state hidden in the 2D transition metal $\rm 2H-MoS_2$ family have only recently been discovered \cite{1,2,3}. They are all protected by space inversion symmetry and have non-zero topological corner charge \cite{10,11,11_1}. The analogous higher-order topological properties can also be extended to 2D ferromagnetic and ferroelectric materials \cite{12,13}. However, there is still a need for more detailed and comprehensive analysis of high-order topological multiorbital systems and phase transition processes, as well as a stronger connection to valleytronics. Moreover, compared to photonic and phononic crystals, which are highly manipulated \cite{14,15}, there are still relatively few electronic materials with higher-order topologies. Therefore, finding new 2D HOTI materials has important value and significance. 

\begin{figure*}
	\centering
	\includegraphics[width=18cm]{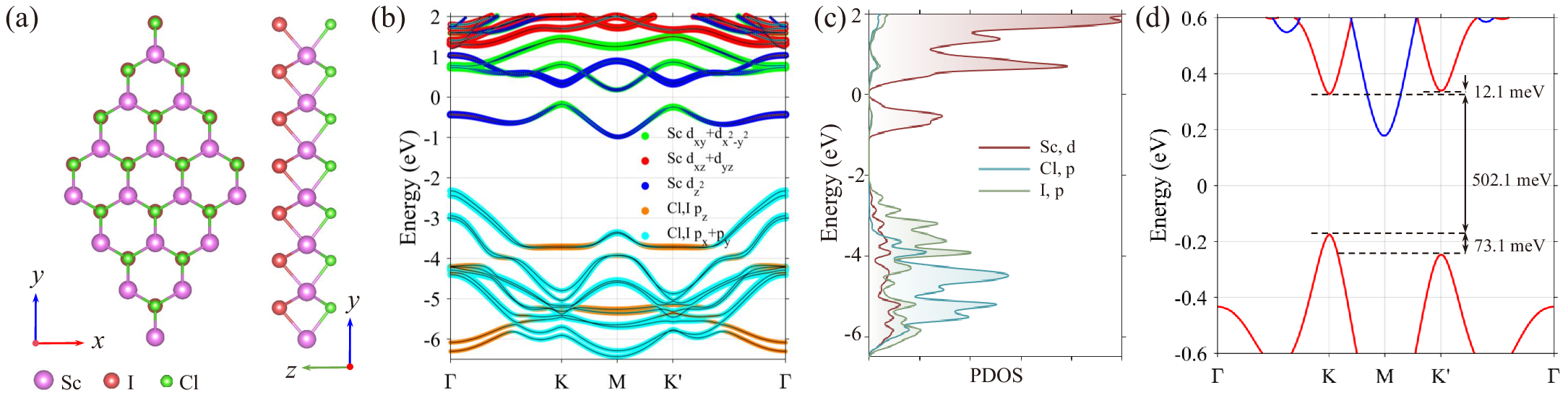}\caption{(a) Top and side views of ScClI. (b) In the out-of-plane ferromagnetic state, a fatband of projected orbitals is present. (c) The projected density of states (PDOS) of the system. (d) Spin-polarized energy bands near the Fermi level. Red and blue lines indicate spin-up and spin-down states.}\label{pho:zak}
\end{figure*}

Two-dimensional materials with honeycomb structure have strong energy valley properties \cite{16,17,18}. For 2D ferromagnetic materials, the intrinsic ferromagnetic order couples with the energy valleys to produce giant valley polarization \cite{4,19} and leads to large differences in the Berry curvature of the K/K' valleys. Since the Berry curvature is analogous to an equivalent magnetic field \cite{20}, the anomalous velocity imparted to the electrons leads to valley currents \cite{18,19,20,21}. Applications on this basis consist of the valley Hall effect \cite{21} and the anomalous Nernst effect \cite{19}.
In particular, the anomalous Nernst effect has been widely measured and applied experimentally \cite{22,23,24,25,26,27,28}. The anomalous Nernst effect on $\rm 2H-MoS_2$ family materials was discussed early by Yu $et \  al.$ \cite{28}. 
It can be predicted that there is a certain connection between the topological properties and the valley Nernst effect. Such a connection has rarely been mentioned. Consequently, associating topological phase transitions with the anomalous valley Hall effect holds significant physical significance.

In this work, we discover that monolayer ScClI is a SOTI, exhibiting second-order topological corner states in a 2D triangular quantum dot. Using the multi-orbital tight-binding approximation theory, we learn that the $p$-orbitals of the halogen elements are indispensable for higher-order topology. It is not possible to show this by only considering the $d$-orbitals. 
Next, by applying external field to change the magneto-valley coupling strength and energy level difference, we provide the complete topological phase diagram of 2D Janus ScClI. ScClI mainly undergoes the topological phase transition process of SOTI, quantum anomalous valley Hall insulators (QAVHI), and normal insulators (NI).
Due to the strong valley electronic properties of ScClI, the phenomena of SOTI, QAVHI, and NI exhibit distinct characteristics in the thermally induced anomalous Nernst effect. 
Based on this foundation, we provide anomalous Nernst conductance maps that correspond to the higher-order topological phase diagram. 
Our results effectively link high-order topology with the anomalous Nernst effect of thermal excitation, which is of great significance for the measurement and application of experimental electronic devices.

\section*{Methods}

First-principles calculations based on density functional theory (DFT) are conducted using the Vienna $ab \ initio$ Simulation Package (VASP) \cite{29,30}. The electronic exchange-correlation interactions are treated using the Perdew-Burke-Ernzerhof (PBE) generalized gradient approximation (GGA) \cite{31,32}. For two-dimensional materials, a minimum of 15 Å vacuum layer is included. The energy cutoff for the plane-wave basis is set to 500 eV, and a $10 \times 10 \times 1$ k-mesh is employed for Brillouin-zone sampling. The convergence threshold for the maximum force during structural optimization is set to be less than 0.01 eV/Å, and the energy convergence criterion is set to be $10^{-6}$ eV. We apply GGA + U method to correct the Coulomb repulsion interactions of the Sc atom's $d$ orbitals, with a typical value of U = 3 eV \cite{33,34}. The parity of the occupied states is calculated using the IrRep program \cite{35}. Additionally, a tight-binding model based on maximally localized Wannier functions (MLWFs) is constructed using the $\rm wannier90$ and $\rm WANNIERTOOLS$ packages \cite{36,37}.

\begin{figure*}
	\centering
	\includegraphics[width=16cm]{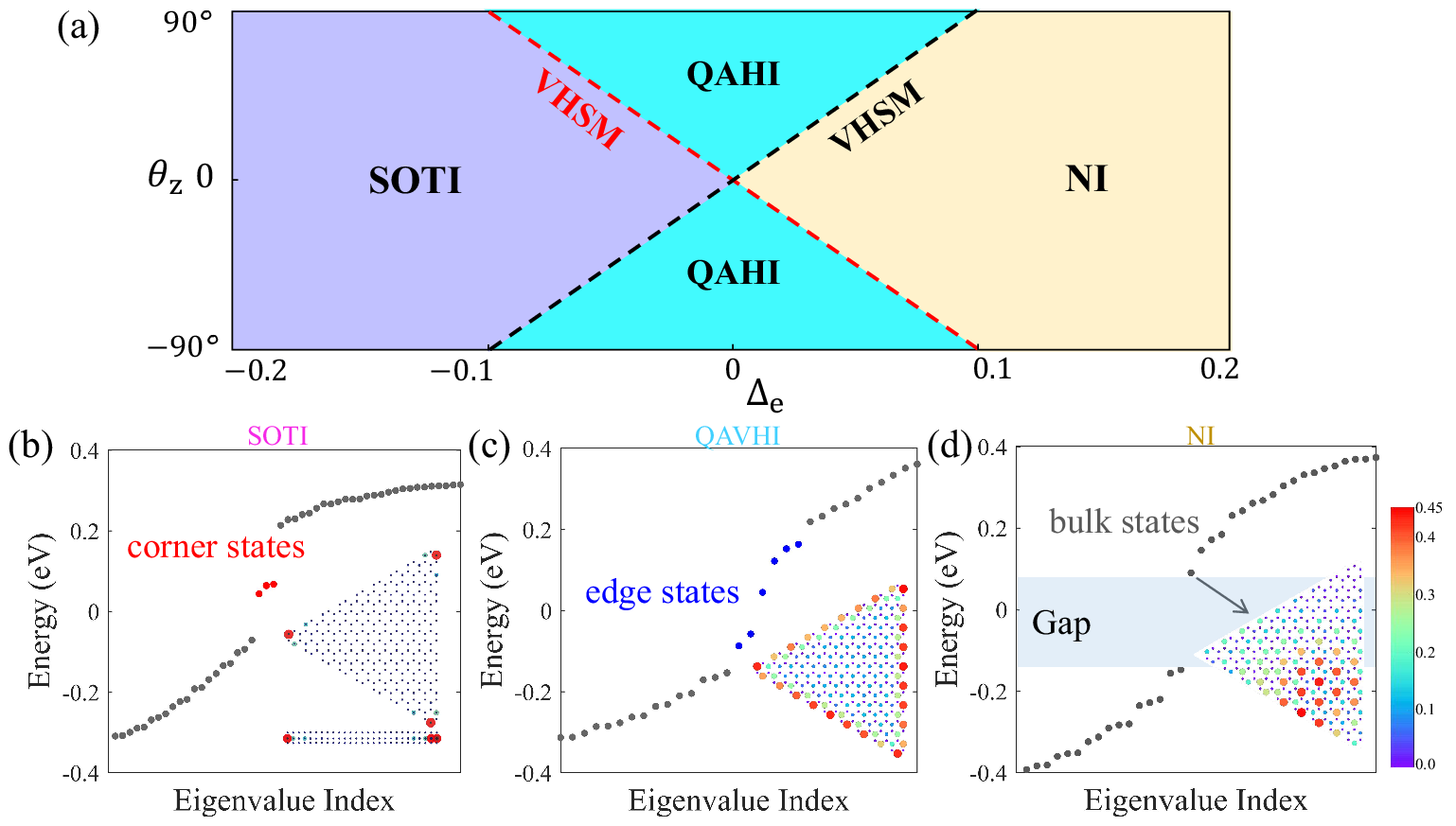}\caption{(a) The complete phase diagram as a function of the magnetic moment direction $\theta_z$ and the energy level difference $\Delta_e$. (b) Energy spectrum of triangular armchair nanosheets. And the total wavefunction distribution in real space for the corner states within the bandgap of the energy spectrum. (c) The quantum dot energy spectrum of QAVHI is displayed, with blue dots marking the boundary states near the Fermi energy. The inset shows the wave function distribution.
		(d) The energy spectrum of NI, illustrating the wave function distribution of bulk states.}\label{pho:phase}
\end{figure*}

\section{ RESULTS AND DISCUSSION}


The crystal structure of the two-dimensional Janus honeycomb material ScClI is shown in Fig. \ref{pho:zak}(a). And its fatband reveals a predominant contribution from the $d$ orbitals of Sc in the vicinity of the Fermi energy level. 
From PDOS, it can be seen that the low energy states dominated by p-orbitals also contain contributions of d-orbitals from Sc, as shown in Fig. \ref{pho:zak}(c).
In the out-of-plane ferromagnetic state, magneto-valley coupling induces polarization at K and K', as illustrated in Fig. \ref{pho:zak}(d). The measured energy level difference between valence electrons in the K and K' valleys is 73.1 meV, and the band gap in the spin-up energy band of the K valley is determined to be 502.1 meV. Although the system is an indirect bandgap insulator in its fully relaxed state, it can be transformed into a direct bandgap insulator by applying strain.

To verify the high-order topological properties of ScClI, we first calculate the higher order topological indices $Q_c^{(3)}$ of the system, which is protected by the symmetry of the $C_3$ rotation symmetry. For the rotation eigenvalues calculation of all occupied states on the high symmetry points in the Brillouin zone, one can take $[\mathrm{K}_n^{(3)}]=\#\mathrm{K}_n^{(3)}-\#\Gamma_n^{(3)}$, where $\#$ denotes the counting about the symmetry eigenvalues at the points K and $\Gamma$. The eigenvalues of the $C_3$ rotations are defined as $e^{2\pi i(n-1)/3}\left(\text{for} \ n=1,2,3\right)$. The topological indices \cite{10,11} of the HOTI are
\begin{equation}
\chi^{(3)}=([K_1^{(3)}], [K_2^{(3)}]),Q_c^{(3)}=\frac{e}{3}[K_2^{(3)}]\text{mod}\ e,
\end{equation}
where $e$ is the charge of the free electron. 
The upper indicator (3) represents $C_3$ symmetry. By performing DFT calculations, we obtain the wave functions of all occupied Bloch states below the Fermi level. Afterwards, utilizing the IrRep program \cite{35} , we calculate the symmetry eigenvalues of these wave functions upon application of the $C_3$ rotation operator. With them, the topological indicator $\chi^{(3)} = (1, 2) $ and the nonzero corner charge $Q_c^{(3)}=2e/3$ is obtained. So, we can conclude that the in-gap corner states of the system is topologically protected.

For this system, the magneto-valley coupling gives rise to a giant valley polarization. This depends on the magnitude of the magnetic moment outside the surface, and the valley polarization disappears when a magnetic moment inside the surface is applied. This is illustrated in Fig. S4 (a,b,c) of the Supplementary Material\cite{sm}.
Then, to comprehensively capture the essential physical characteristics of higher-order topology, we develop a simplified multi-orbit tight-binding (TB) model based on the Wannier functions. 
To provide a complete representation of HOTI, the TB model incorporates the entire set of five $d$-orbitals and six $p$-orbitals (from Sc, Cl and I, respectively), with a focus on the nearest-neighbor orbital hopping. 
From Wannier fitting, the system shows both spin-orbit coupling (SOC) and valley polarization due to intrinsic magnetism. Thus, the Hamiltonian can be written as:

\begin{equation}
H=\sum_{\langle i, j\rangle\alpha,\beta}(t_{i j}^{\alpha,\beta}c_{i\alpha}^{\dagger}c_{j\beta}+h.c.)+t_{s o c}\bm{L}\cdot \bm{S}+m_{z}\bm{M}\cdot \bm{S},
\end{equation}
where $c_{i\alpha}^{\dagger}(c_{j\beta})$ represents the electron creation (annihilation) operator for the orbital $\alpha(\beta)$ located at position $i(j)$. $t_{soc}$ denotes the strength of the SOC, and $M = (M_x, M_y, M_z)$ represents the magnetic moment direction with $m_z$ intensity. $S$ is the Pauli matrix.

Under this TB model, we have successfully identified higher-order topological corner states in the fully relaxed structure of ScClI. The electronic band structure of the triangular quantum dot is depicted in Fig. \ref{pho:phase}(b), where the presence of red color in the middle of the band gap represents the corner states. 
The wave functions of the corner states are localized on the corners of the quantum dots. Higher order topological corner states are the result after full consideration of the $p, d$ orbitals.
Conversely, if we focus solely on the $d$-orbital contribution from Sc atom,  the presence of only one occupied band below the Fermi level results in the corner charge $Q_c^{(3)}=0$. In this way, we will not obtain in-gap corner states that are topologically protected, as shown in Fig. S2. Moreover, from the PDOS, it can be seen that there is a certain coupling between the $d$ and $p$ orbitals. So, establishing a TB model for multiple orbits is effective. 

Next, we can achieve higher-order topological phase transitions by tuning the angle $\theta_z$ of the magnetic moment and the energy level difference $\Delta_e$ between the $d_{xy}$ and $d_{z^2}$ orbitals at the K-valley vertex. The angle $\theta_z$ primarily influences the magnitude of valley polarization, while $\Delta_e$ induces band inversions, leading to a transition from a SOTI to a QAVHI. Due to the magneto-valley coupling, after the first band inversion, the system becomes a QAVHI with Chern number $C=1$. Following the second band inversion, the system transforms into a trivial insulator. Notably, at the critical point of the topological phase transition, the system manifests a valley-hall semimetal phase at the K valley during the initial inversion and at the K' valley during the subsequent inversion. Fig. \ref{pho:phase}(a) displays the complete phase diagram, showing the progression SOTI-VHSM-QAVHI-VHSM-NI.

\begin{figure*}
	\centering
	\includegraphics[width=14cm]{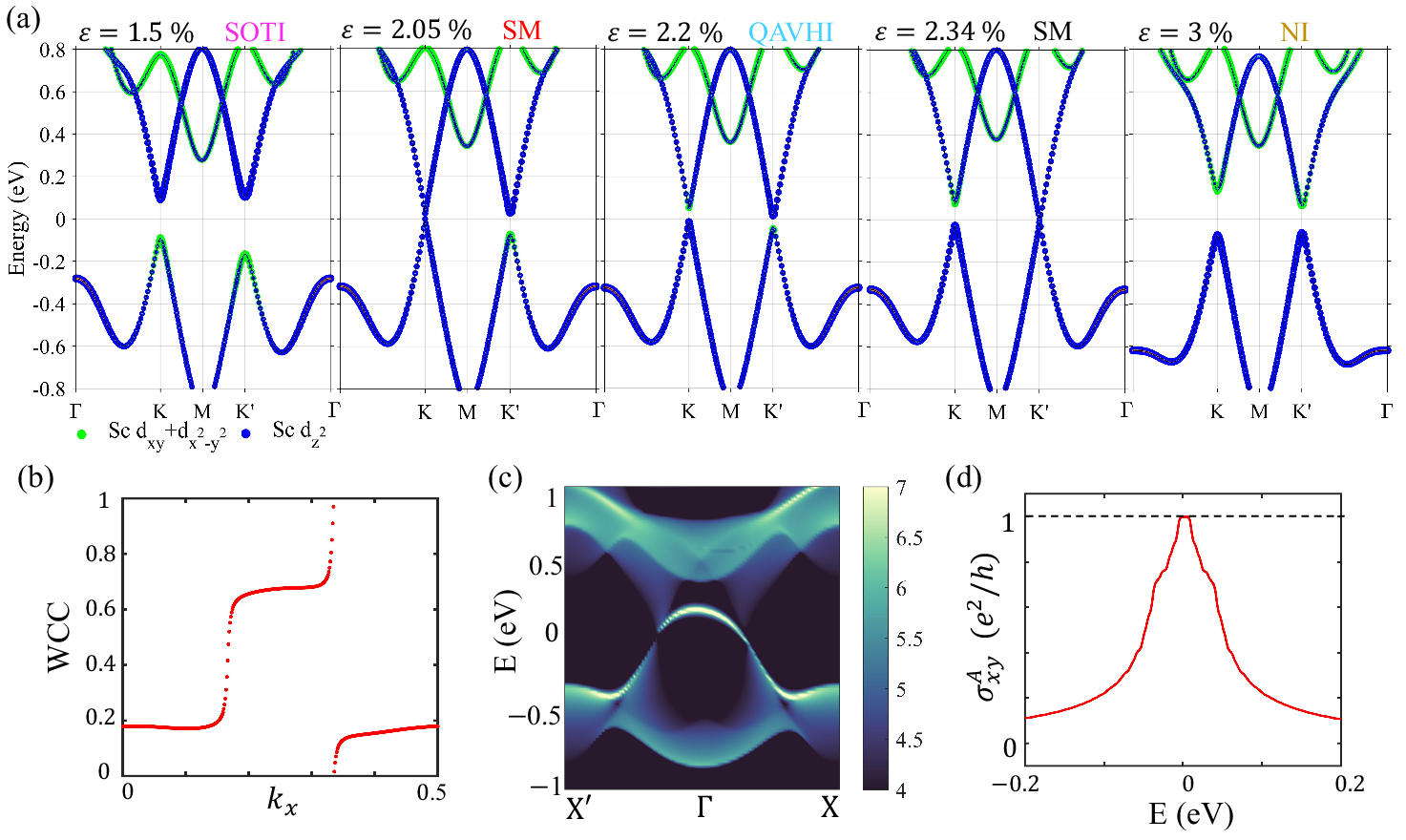}\caption{ 
		(a) Projected energy bands evolution in the $1.5-3\%$ strain range. 
		WCC and surface band of QAVHI at $2.2\%$ strain are shown in (b) and (c), respectively. (d) Plot of quantum anomalous Hall conductance versus energy.}\label{pho:QAVHI}
\end{figure*}


For the above topological phase transition process, we can realize it by applying tensile strain. After experiencing the first energy band inversion in the K valley, the system becomes QAVHI. At this time, the energy spectrum of the triangular quantum dots undergoes a transformation, wherein the band gap disappears and is supplanted by a continuum of edge states, as shown in Fig. \ref{pho:phase}(c). With increasing strain, the second energy band inversion occurs in the K' valley. The system behaves as a normal insulator. The corresponding energy spectrum exhibits a band gap with no corner states or edge states within it. Additionally, both sides of the bandgap show localized features within the body, as shown in Fig. \ref{pho:phase}(d). At tensile strain magnitudes of $2.05\%$ and $2.34\%$, the system reaches the phase boundary, transitioning into a Valley-half Semi-metal, as shown in Fig. \ref{pho:QAVHI}(a). In both strain ranges, the Chern number of the system is 1, and Fig. \ref{pho:QAVHI}(b) illustrates the Wannier charge center (WCC) of the system. The surface states and quantum anomalous Hall conductance platform of QAVHI are shown in Fig. \ref{pho:QAVHI}(c, d), respectively.

\begin{figure*}
	\centering
	\includegraphics[width=12cm]{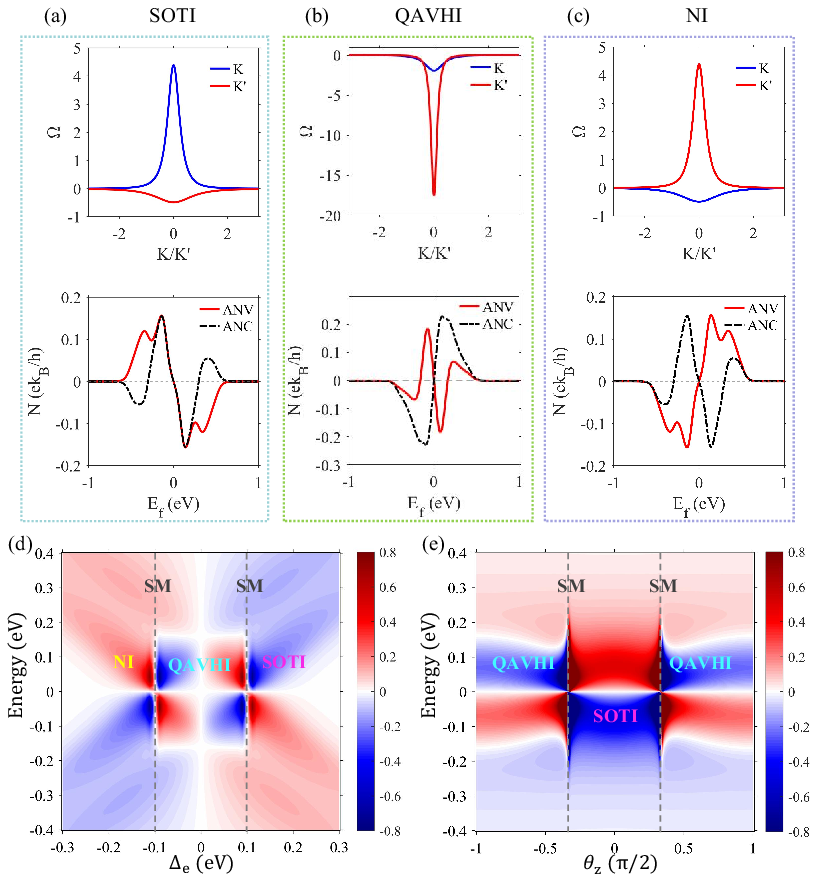}\caption{(a), (b), and (c) present the Berry curvature, ANC and ANV of three insulation correlations: SOTI, QAVHI, and NI, respectively, concerning the K/K' valley.
		In (d), ANV is plotted as a function of energy and gap. Meanwhile, (e) illustrates the relationship between ANV, energy, and the direction of the magnetic moment $\theta_z$. Phase boundaries are marked by gray dashed lines.}\label{pho:ANE}
\end{figure*}


The giant valley polarization property, due to the intrinsic magnetism of the system, has attracted our attention because it will bring about a difference in the Berry curvature at the valley of K/K'. Berry curvature is the same as adding an equivalent magnetic field to the system, which causes the electrons to acquire anomalous velocities \cite{20}. We can exploit this property to obtain tunable valley flows.

Since the Berry curvature induced valley flow depends on the neighborhood of the Fermi energy level and related to the valley degree of freedom, we can use the two-band k-p model to characterize the physical properties of valley locking. The Hamiltonian of the two-band can be written as:
\begin{equation}
\mathcal{H}=at(k_x\sigma_x+\eta k_y\sigma_y)+\frac{\Delta_e}{2}\sigma_z+ \sin{\theta_z} \eta m_z\sigma_z,
\end{equation}
where is $\eta$ the valley index, $a$ is the lattice constant, $t$ is the hopping integral. 
$m_z$ denotes the strength of the magneto-valley coupling. $\sigma_{x,y,z}$ is the Pauli matrix. The absence of a spin indicator is attributed to the spin polarization in the vicinity of the Fermi energy level.  Effective mass $ \Delta_m = \Delta_e /2 + \sin{\theta_z} \eta m_z $. The energy eigenvalues are
\begin{equation}
E_\eta^n=n\sqrt{\left(kat\right)^2+{\Delta_m}^2},
\end{equation}
where $n = \pm 1$, denoting the conduction and valence bands, respectively. According to the definition of Berry curvature $\Omega(k)=\nabla\times\langle u(\textbf{k})|i\textbf{V}_i|u(\textbf{k})\rangle$, where $u(k)$ is the periodic part of the Bloch wave function. We can obtain the Berry curvature associated with the valley
\begin{equation}
\Omega_\eta^n(\mathbf{k})=-\eta n\frac{a^2t^2\Delta_m}{2\left[\left(k at\right)^2+\left(\Delta_m\right)^2\right]^{(3/2)}}
\end{equation}
By tuning the bandgap and the magneto-valley coupling strength, we obtained the Berry curvature for the three phases SOTI, QAVHI and NI, respectively. They are shown in Fig. \ref{pho:ANE}(a-c), respectively, matching the results of the DFT calculations. It is worth noting that the valley polarization is reflected by the difference in Berry curvature. One of the distinguishable traits of the Quantum Anomalous Valley Hall Effect (QAVHI) is the evident shift from opposite polarities to consistent polarities in the Berry curvature of the K/K' valleys during a phase transition.

For characterizing the nature of the valleys matching the phase diagram, we can reveal it through the anomalous Nernst effect of thermal excitation. The valley-dependent anomalous Nernst conductivities is defined as:
\begin{equation}
\mathcal{N}_\eta=\frac{ek_{B}}{\hbar}\sum_n\int\frac{d^2k}{{(2\pi)}^2}\Omega_\eta^n\mathcal{S}_\eta^n(\mathbf{k}),
\end{equation}
where $\mathcal{S}_\eta^n$ is the entropy density, $f_k$ denotes the Fermi distribution function. At 300 K temperature, we can obtain the anomalous valley Nernst conductance (ANV) and the anomalous charge Nernst conductance (ANC):
\begin{equation}
\text{ANV}=\mathcal{N}_k-\mathcal{N}_{k'}, \ \text{ANC}=\mathcal{N}_k+\mathcal{N}_{k'},
\end{equation}

Figure \ref{pho:ANE}(a-c) shows the anomalous Nernst conductivities for each of the three phases. In the context of QAVHI, the ANC intersects the zero energy level just once, whereas for SOTI and NI, the ANC crosses the zero point three times. Therefore, the ANC does not perfectly distinguish between individual insulation phases. On the other hand, the ANVs of the three phases differ from each other. The ANV of QAVHI has three intersections with the Fermi energy level. The ANVs of SOTI and NI has only one intersection with the Fermi energy level, and they have opposite signs. Therefore, ANV is a perfect choice to distinguish the insulating phases. 
When the magnetic moment is perpendicular to the plane, the magneto-valley coupling strength reaches its maximum. By adjusting the gap size, the contour plot of the ANV with respect to energy and gap is obtained, as shown in Fig. \ref{pho:ANE}(d), with this process accompanied by a fixed magnetic moment direction $\theta_{z} = 90^{\circ}$.  
Fig. \ref{pho:ANE}(e) corresponds to a fixed energy level difference of $\Delta_e = -0.05$ eV, with magnetic moment direction adjustment, resulting in a QAVHI-SOTI-QAVHI phase transition that perfectly matches the phase diagram.

\section{CONCLUSIONS}

In summary, we employ the multi-orbital tight-binding method to analyze the high-order topological properties of the 2D Janus ScClI. We give the complete phase diagram of ScClI's high-order topology, which is controlled by the magnetic moment direction $\theta_z$ and the orbital energy level interpolation $\Delta_e$. The theoretical predicted phase transition process of SOTI-VHSM-QAVHI-VHSM-NI has been verified and realized from the DFT calculation results under strain engineering. 
Meanwhile, due to the valley polarization characteristics of the system, the three types of insulators, SOTI, QAVHI, and NI, have different anomalous Nernst conductivity. By exploiting the anomalous Nernst effect, we construct an ANV map that corresponds to the topological phase diagram.
Our findings can be extended to other same type of two-dimensional materials, paving a way for the future measurements and characterizations of higher-order topological phases in a broader range of materials.

\section*{ACKNOWLEDGMENTS}

We acknowledge the financial support by the National Natural Science Foundation of China (No. 11874113) and the Natural Science Foundation of Fujian Province of China (No. 2020J02018).

\bibliography{References.bib}


\end{document}